\documentclass[prl,twocolumn,preprintnumbers,amsmath,amssymb]{revtex4-1}
\usepackage{graphicx}

\begin{document}
\title{ Topological depletions and universal sub-leading scalings across topological phase transitions   }
\author{ Fadi Sun  and Jinwu Ye  }
\affiliation{   Department of Physics and Astronomy, Mississippi State  University, P. O. Box 5167, Mississippi State, MS, 39762   \\
Department of Physics, Capital Normal University,
Key Laboratory of Terahertz Optoelectronics, Ministry of Education, and Beijing Advanced innovation Center for Imaging Technology,
Beijing, 100048, China  \\   
Kavli Institute of Theoretical Physics, University of California, Santa Barbara, Santa Barbara, CA 93106 }
\date{\today }

% Near the $ U(1) $ limit, $ \beta $ is small.
% Near the $ Z_2 $ limit, $ 1-\beta $ is small.
% The combination of the two methods provides rather complete physical picture from the $ U(1) $  to $ Z_2 $ limits.

\begin{abstract}
%   It was well established that physical quantities satisfy universal scaling functions across a quantum phase transition with an order parameter.
%   Similar scaling functions may also hold across a topological quantum phase transition (TPT) without an order parameter scaling to a single
%   point in momentum space.
   It remains an open problem if there are universal scaling functions across a topological quantum phase transition (TPT)  without an order parameter, but
   with extended Fermi surfaces (FS ).
   Here, we study a simple system of fermions hopping in a cubic lattice subject to a Weyl type spin-orbit coupling (SOC).
   As one tunes the SOC parameter at the half filling, the system displays Weyl fermions and
   also various TPT due to the collision of particle-particle or hole-hole Weyl Fermi Surface (WFS).
   At the zero temperature, the TPT is found to be a third order one whose critical exponent is determined.
 We derive the scaling functions of the specific heat, compressibility and magnetic susceptibilities.
 In contrast to all the previous cases in quantum or topological transitions,
 although the leading terms are non-universal and cutoff dependent, the sub-leading terms satisfy universal scaling relations.
% This fact is a unique and salient feature of this kind of TPT with extended FS reconstructions.
 The sub-leading scaling leads to the topological depletions (TD) which show  non-analytic and non-Fermi liquid corrections
 in the quantum critical regime, can be easily distinguished from the analytic leading terms and detected experimentally.
%   Away from half filing, we determine the global phase diagram of the chemical potential versus the SOC parameter.
%   the transition is driven by particle WFS or hole WFS with the half filling case as the Tetra-critical point.
 One can also form a topological Wilson ratio from the TD of two conserved quantities
 such as the specific heat and the compressibility. As a byproduct, we also find Type II Weyl fermions appearing as
 the TPT due to the collision of the extended particle-hole WFS.
 Experimental realizations and detections in cold atom systems and materials with SOC are discussed.
%    we determine the global phase diagram of the chemical potential versus the SOC parameter.
%   the transition is driven by particle WFS or hole WFS with the half filling case as the Tetra-critical point.
%   In contrast to most other quantum or topological transitions, only sub-leading terms satisfy universal scaling relations,
%   while leading terms are non-universal and cutoff dependent.
% This may be the first work to combine scaling concepts in quantum critical phenomena with a topological transition where
% symmetry breaking and order parameters are absent. This combination may lead to a new universality class of scaling functions: local QCP
% which is unique to fermionic  Lifshitz type of TQPT with extended FS.
\end{abstract}

\maketitle

{\bf 1. Introduction. }
% Quantum phase transitions with an order parameter have been under intense investigations
% since the experimental  discovery of high temperature superconductivity.
 It was established that experimental measurable quantities near a quantum phase transition with an order parameter
 satisfy various universal scaling functions at a finite temperature \cite{scaling,sachdev}.
 In an other forefront, topological phases and phase transitions without an order parameter
 were studied since the experimental observations of the quantum Hall effects \cite{wenniu,volovik}
 and under even more intense investigations after the discovery of  topological insulators \cite{kane}.
% Topological phenomena in various fermionic systems \cite{volovik} have been revived since the more recent
% experimental realizations of a new kinds of insulators called topological insulators \cite{kane,zhang}.
 It is important to investigate if there are still universal scaling functions across various topological phase transitions without
 an order parameter.
% There are previous efforts to derive leading scaling functions across a TPT such as the Quantum Hall
% to an insulator transition in \cite{dirac1,dirac3} and that driven by collisions
% of Dirac points in a honeycomb lattice \cite{tqpt}. All these TPT  can be scaled to a single point in momentum space,
% so conventional Renormalization group (RG), large $ N $ expansion or other methods can be applied to capture
% low energy critical fluctuations and derive the scaling functions.
 Here, we address this outstanding problem by studying a very simple system of free fermions hopping in a
 cubic lattice subject to a Weyl type of spin-orbit coupling \cite{socsdw}. Our main results are summarized in the abstract.
 The experimental motivations of this model from both cold atoms and materials will be discussed in Sec.7.

\begin{figure}
 \includegraphics[width=2.75cm]{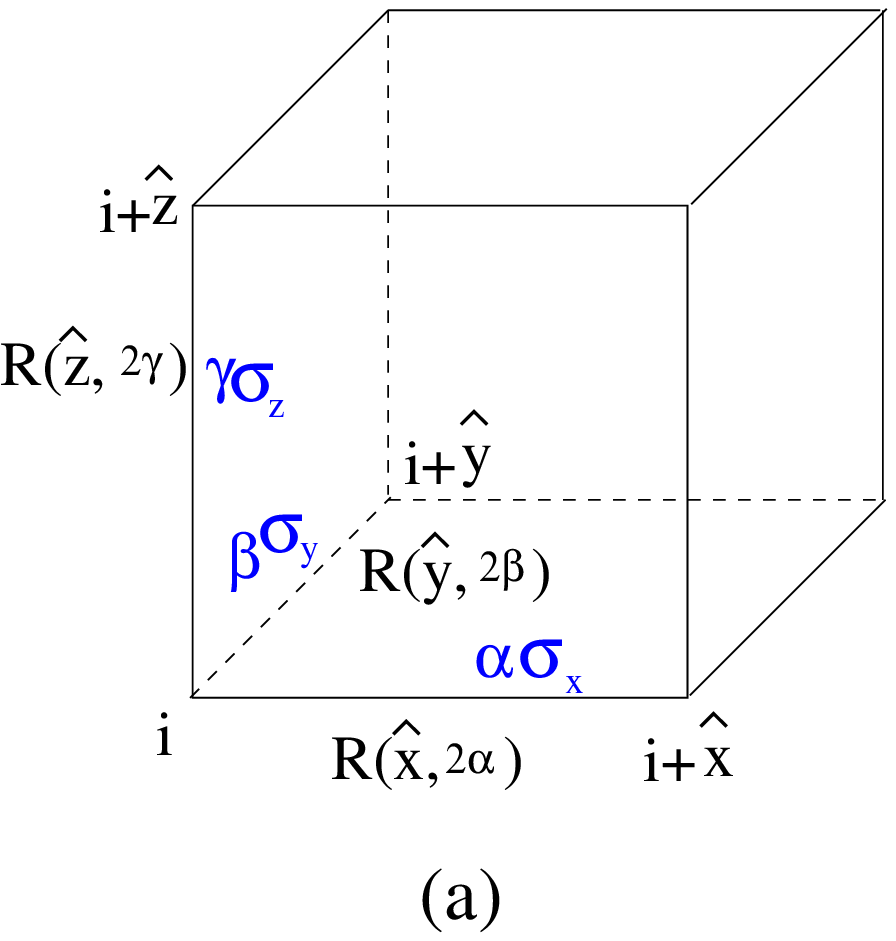}
%\hspace{0.15cm}
%\includegraphics[width=2.75cm]{cubenonb.eps}
\hspace{0.25cm}
\includegraphics[width=2.75cm]{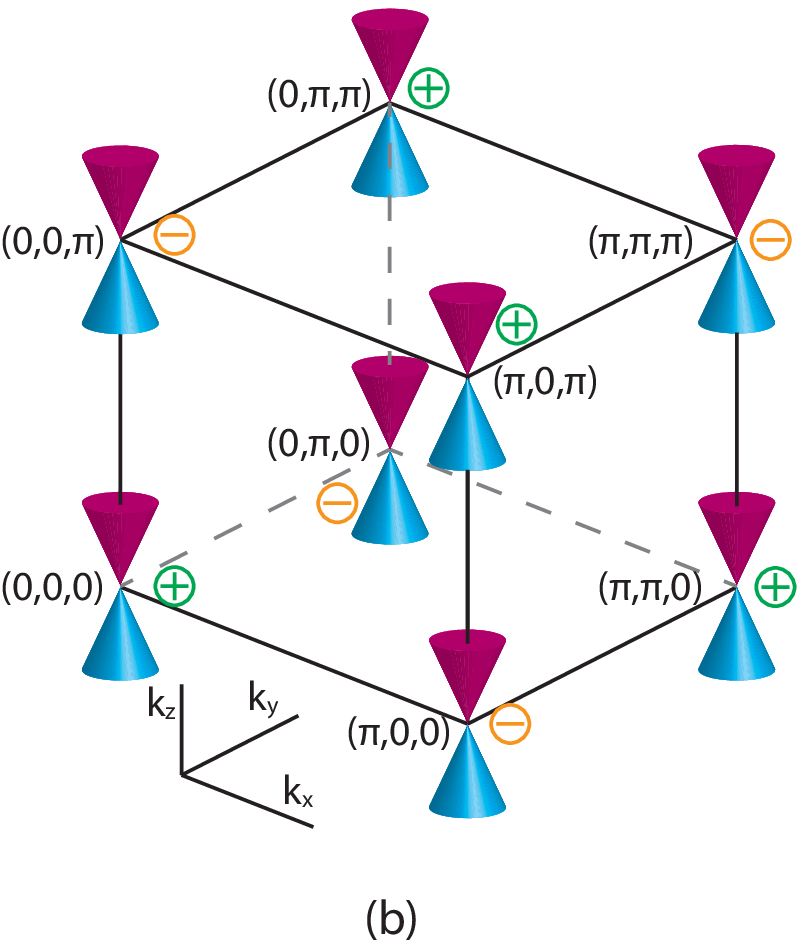}
%\hspace{0.5cm}
%\includegraphics[width=3cm]{fmcritical.eps}
\caption{ (a) The non-abelian gauge fields $ ( \alpha \sigma_x, \beta \sigma_y, \gamma \sigma_z ) $ are put on the three links in a cubic lattice.
%(b) The $ ( \alpha,\beta,\gamma ) $ SOC parameter space.
%    There are one $ SU(2) $ Abelian point at the origin and three more different Abelian points in the
%    correspondingly rotated frames $ \widetilde{SU}(2), \widetilde{\widetilde{SU}}(2), \widetilde{\widetilde{\widetilde{SU}}}(2) $
%    at the edge, face and the cubic center respectively.
%    There are Type I Weyl fermions along the line connecting the cubic center to the edge center and
%    Type II Weyl fermions at the face center.
(b) The 8 Weyl fermions with the topological charges $ N_3=\pm 1 $ at $ \alpha=\beta=\gamma=\pi/2 $. }
%    See Fig.\ref{type1} for the Weyl point explanation. }
\label{cubicweyl}
%    It was established \cite{largecluster} that at the two $ SU(2) $ points at $ \theta=0 $ and $ \theta= \pi $,
%    there is a direct second order phase transition from the semi-metal phases to the AFM state at $ U/t \sim 3.8 $,
%    no intervening  gapped $ Z_2 $ spin liquid phase as claimed in \cite{meng}. However, it is not known if there is an intermediate $ Z_2 $ spin liquid
%    between the semi-metal phases and the $ Y-x $ stripy state away from the two $ SU(2) $ points.
\end{figure}

{\bf 2. 8 Type I Weyl fermions. }
  The Hamiltonian  of fermions hopping in a cubic lattice subject to Weyl type spin-orbit coupling in Fig.\ref{cubicweyl}a can be written as
\begin{equation}
  H= \sum_k h_i( \boldsymbol{k} ) \sigma_i,i=0,1,2,3
\label{ham}
\end{equation}
where $ \sigma_i= \sigma_0 $ is the identity matrix, $ \sigma_i $ are 3 Pauli matrices and
$ h_0(k)=-2t(\cos\alpha\cos k_x+\cos\beta\cos k_y+\cos\gamma\cos k_z), h_x(k)=2t\sin\alpha\sin k_x,
h_y(k)=2t\sin\beta\sin k_y, h_z(k)=2t\sin\gamma\sin k_z $.
%The Non-abelian gauge parameters $ ( \alpha,\beta,\gamma ) $ are shown in Fig.\ref{cubicweyl}b.
Its two energy bands are $ \epsilon_\pm(k)=h_0(k)\pm h(k) $ where $ h(k)=\sqrt{[h_x(k)]^2+[h_y(k)]^2+[h_z(k)]^2} $.
At half filling $ \mu=0 $, the particle and hole FS is given by $
  \epsilon_\pm(k)=0  $. The particle energy is related to that of the hole $ \epsilon_{+}( \vec{k} + \vec{Q} )= - \epsilon_{-}(\vec{k} ) $
 where $ \vec{Q}=( \pi, \pi, \pi) $ is the FS nesting vector which  separates the particle FS from the hole FS.
% It is this separation which distinguishes the TPT in particle-particle or hole-hole from that of Type I in
% Fig.\ref{type1} and Type II Weyl fermions in Fig.\ref{type2} where the particle WFS collides with the hole WFS.
 It leads to the relation between the particle DOS and that of hole $ D_{+}( \omega )= D_{-} ( - \omega ) $ at the half filling $ \mu=0 $.

\begin{figure}
\includegraphics[width=8cm]{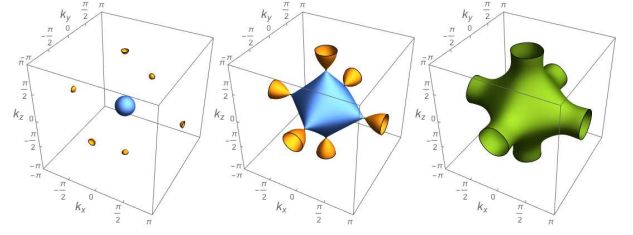}
%\hspace{0.5cm}
%\includegraphics[width=3cm]{fmcritical.eps}
\caption{
 The particle WFS evolves along $ \alpha=\beta=\gamma=\theta $ for $ \theta=4\pi/9,\pi/3, \pi/4 $.
%When slightly away from the center  $ \delta \theta=\pi/2-\theta \ll 1 $ with $ \delta \theta \ll 1 $,
%The Weyl fermion ( blue ) at $ (0,0,0) $ with $ N_3=1 $ receives a finite chemical potential $ 3 \mu_0 >0 $ ( particle-like )
%where $ \mu_0=2 t \delta \theta $. The Weyl point at $ (\pi,\pi,\pi) $ with  $ N_3=-1 $ receives a
%finite chemical potential $ -3 \mu_0 $ ( hole-like),
%while the other 6 Weyl fermions with $ N_3=\pm 1 $ receive a finite chemical potential $ \mu_0, - \mu_0 $ respectively.
At the QCP $ \theta_c=\pi/3 $, the big WFS with $ N_3=1 $ (blue)  hits the other 3 small WFS ( yellow )
with $ N_3=-1 $ at 6 Fermi points at $ \vec{K}_c= (\pm 2\pi/3,0,0), (0,\pm 2\pi/3,0), (0,0,\pm 2\pi/3 ) $. As $ \theta $ increases further,
it becomes a whole ( green faucet tap-like ) Fermi surface  with the total $ N_3=1-1-1-1=-2 $ charge, so it is a topological non-trivial FS.
The hole WFS can be reached by shifting the particle WFS by the FS nesting vector $ ( \pi,\pi,\pi) $. }
\label{tqpt1}
%    It was established \cite{largecluster} that at the two $ SU(2) $ points at $ \theta=0 $ and $ \theta= \pi $,
%    there is a direct second order phase transition from the semi-metal phases to the AFM state at $ U/t \sim 3.8 $,
%    no intervening  gapped $ Z_2 $ spin liquid phase as claimed in \cite{meng}. However, it is not known if there is an intermediate $ Z_2 $ spin liquid
%    between the semi-metal phases and the $ Y-x $ stripy state away from the two $ SU(2) $ points.
\end{figure}

\begin{figure}
\includegraphics[width=8cm]{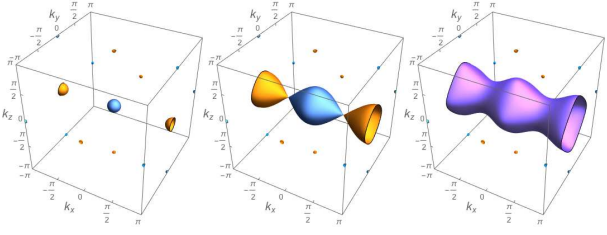}
%\hspace{0.5cm}
%\includegraphics[width=3cm]{fmcritical.eps}
\caption{  The particle WFS evolves along $ \alpha=\pi/2, \beta=\gamma=\theta $ for $ \theta=4\pi/9,\pi/3, \pi/4 $.
%  The end point $ \theta=0 $ ( not shown ) is the $ 0 $ flux ( in all the three planes )  Abelian point
%  with the $ \tilde{SU}(2) $ symmetry in the rotated basis in Fig.\ref{cubicweyl}b.
%        When away from the center  $ \delta \theta=\pi/2-\theta \ll 1 $ with $ \theta \ll 1 $,
%the Weyl fermion at $ (0,0,0) $ with $ N_3=1 $ and the one at $(\pi,0,0) $ with $ N_3=-1 $
%receive a finite chemical potential $ 2 \mu_0 >0 $ ( particle-like )  where $ \mu_0=2 t \delta \theta $.
%Similarly, the Weyl fermion at $ (0,\pi,\pi ) $ with $ N_3=1 $ and the one at $ (\pi,\pi,\pi) $ with $ N_3=-1 $
%receive $ -2 \mu_0 < 0  $ ( hole-like ).
%The other 4 Weyl fermions with $ N_3 =\pm 1 $ remain intact.
At $ \theta_c=\pi/3 $, the WFS with $ N_3=1 $ hits the one with $ N_3=-1 $ at the two
Fermi points at $ \vec{K}_c=(\pm \pi/2,0,0) $. As $ \theta $ decreases further,
it becomes a whole Fermi surface ( violet vase ) with the total $ N_3=1-1=0 $ charge, so it is topologically trivial FS.
There are 4 Weyl fermions remaining intact through the TPT.
%This should be in a  different class of TPT than in Fig.\ref{tqpt1}.
%The co-existence of the 4 Type I Weyl fermions and the TPT of WFS is one of the new features along this line.
The hole WFS can be reached by shifting the particle WFS by one of the two FS nesting vectors
$ ( 0,\pi,\pi), ( \pi,\pi,\pi) $.}
\label{tqpt2}
%This should be a different class of TPT than {\bf 1} above. }
%    It was established \cite{largecluster} that at the two $ SU(2) $ points at $ \theta=0 $ and $ \theta= \pi $,
%    there is a direct second order phase transition from the semi-metal phases to the AFM state at $ U/t \sim 3.8 $,
%    no intervening  gapped $ Z_2 $ spin liquid phase as claimed in \cite{meng}. However, it is not known if there is an intermediate $ Z_2 $ spin liquid
%    between the semi-metal phases and the $ Y-x $ stripy state away from the two $ SU(2) $ points.
\end{figure}

\begin{figure}
%\includegraphics[width=3.75cm]{cubicbosonu1.eps}
%\hspace{0.25cm}
\includegraphics[width=8cm]{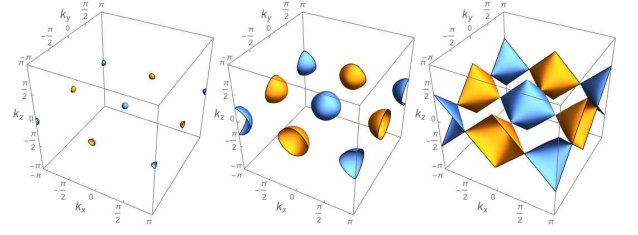}
%\hspace{0.5cm}
%\includegraphics[width=3cm]{fmcritical.eps}
\caption{
%(a) The corresponding $ U(1) $ basis at the solvable line $ \alpha=\beta=\pi/2, \gamma=\theta $ with $ \sigma_z=\pm 1 $.
%Compare with the $ U(1) $ basis for a square lattice in Fig.1b and a honeycomb lattice in Fig.7a2.
        The particle WFS evolves along   $ \alpha=\beta=\pi/2, \gamma=\theta $ for $ \theta=4\pi/9,\pi/4, 0 $.
        The TPT happens at the end point $ \theta_c=0 $ where it becomes a corner-sharing octahedron.
        The hole WFS can be reached by shifting the particle WFS by one of the 4 FS nesting vectors
        $ (0,0,\pi), (\pi,0,\pi), ( 0,\pi,\pi),(\pi,\pi,\pi) $.
        There are also 8 Type II Weyl fermions at $ \theta_c=0 $ due to the collision of the particle WFS  and the hole WFS. }
\label{tqpt3}
%    It was established \cite{largecluster} that at the two $ SU(2) $ points at $ \theta=0 $ and $ \theta= \pi $,
%    there is a direct second order phase transition from the semi-metal phases to the AFM state at $ U/t \sim 3.8 $,
%    no intervening  gapped $ Z_2 $ spin liquid phase as claimed in \cite{meng}. However, it is not known if there is an intermediate $ Z_2 $ spin liquid
%    between the semi-metal phases and the $ Y-x $ stripy state away from the two $ SU(2) $ points.
\end{figure}

%The $ (\alpha, 0, 0) $ line is just Abelian point.
%The  blue dashed line $ (\alpha=\pi/2, \beta, 0) $  is just a trivial 3d extension of the 2d case studied in \cite{rh}.
%The green dashed line $ (\alpha, \alpha, 0) $ is also a trivial 3d extension of the 2d case studied in Part II.
At $ ( \alpha,\beta,\gamma )= (\pi/2,\pi/2,\pi/2 ) $,
there are 8 Type I Weyl fermions located at $ ( k_x=0,\pi, k_y=0,\pi, k_z=0,\pi ) $
carrying the topological monopole charges $ N_3= \pm 1 $ in Fig.\ref{cubicweyl}b.
%The center is the $ \pi $ flux ( in all the three planes ) Abelian point with the $ \tilde{\tilde{\tilde{SU}}}(2) $ symmetry in the rotated basis.
It is the inversion symmetry breaking in Eq.\ref{ham} which leads to their existences.
%Its dispersion relation $ \epsilon^{I}_{\pm}(\vec{q})=\pm \sqrt{ q^2_x + q^2_y + q^2_z } $ leads to the dynamic exponent $ z=1 $
%and a vanishing DOE $ D(\omega) \sim \omega^2 $, so it is a semi-metal. Conventional scaling analysis with $ z=1 $ in \cite{scaling,sachdev}
%leads to $ C_v \sim T^{3}, \chi_u \sim T^{2} $ which can be easily distinguished from convention Fermi Liquid (FL) behaviors $ C_v \sim T, \chi_u \sim C $.
%Here we focus on topological phase transitions (TPT) along the three lines emanating from the 8 Type I Weyl fermions at the half filling $ \mu=0 $.
%These phenomena are unique to 3d SOC systems and can not be considered as a layered structure of a 2d SOC systems
%( just like a strong 3d TI can not be considered as a layered structure of a 2d TI ).
%We will only focus on the half filling case $ \mu=0 $.
As one tunes the SOC parameters, some or all Weyl fermions will become closed particle or hole Weyl Fermi surface (WFS) which
still keep the topological charges $ N_3=\pm 1 $ of the Weyl fermions and
satisfy $ \sum^{8}_{i=1} N_{3i}=0 $ during the evolution \cite{volovik}.
How the WFS evolve along the  three lines $ (\alpha=\beta=\gamma=\theta), (\alpha=\pi/2, \beta=\gamma=\theta) $
and $( \alpha=\beta=\pi/2, \gamma=\theta ) $ are shown in Fig.\ref{tqpt1},\ref{tqpt2},\ref{tqpt3} respectively.
%The WFS satisfy $ \sum^{8}_{i=1} N_{3i}=0 $ during the evolution.
%Extend our analysis to the doping cases with $ \mu \neq 0 $ and map out the global topological phase diagrams in
%the chemical potential $ \mu $ and the SOC parameter space will be presented in subsequent works.

%Following the strategies outlined in the previous sections, starting the solvable line in Fig.8, we will be able to understand
%the physics at other parameters $ (\alpha,\beta,\gamma) $, especially along the two line in Fig.9,10.
%We will work out all the new physics unique to the 3d Weyl fermions at both strong or weak $ U > 0 $ or $ U < 0 $, especially the conditions to
%generate either TR invariant TSF at $ U >0 $ away from half filling or chiral gapless Weyl TSF at $ U <0 $ in a Zeeman field.

% We expect a Van Hove singularity at the TPT $ \theta_c=\pi/3 $. The nature of the TPT in Fig.9 remain to be determined.

{\bf 3. The third order TPT  along $ \alpha=\beta=\gamma=\theta $  at zero temperature.  }
% In this section, we focus on along the diagonal line $ \alpha=\beta=\gamma=\theta $ in Fig.\ref{cubicweyl}b
% and at the half filling $ \mu=0 $.
%How the WFS evolves along  the diagonal line $ \alpha=\beta=\gamma=\theta $ is shown in Fig.\ref{tqpt1}.
There is a TPT driven by the collisions of the 4 WFS where
the colliding 4 particle WFS takes a saddle point ( cone ) geometry near a Von Hove singularity  $ K_c=2 \pi/3 $
and the critical SOC parameter  $ \theta_c=\pi/3 $. When expanding around the VHS $ K= K_c+\Delta/\sqrt{3} $ and
$ \theta_c $, we get the particle energy spectrum:
\begin{equation}
   \epsilon_{+}( \vec{q} )=-[ \Delta + a q^{2}_x-b (q^{2}_y + q^{2}_z) ]
\label{TPTdis}
\end{equation}
   where $ \vec{k}= \vec{K}+ \vec{q}, \Delta=\sqrt{3}( \theta_c-\theta ) $ is the tuning parameter
   and $ a=1/2, b=3/4 + \Delta/4 $.

   The DOS takes the piece-wise form:
\begin{eqnarray}
    D(\omega) = \left \{ \begin{array}{ll}
	 B[ \Lambda - \sqrt{ \frac{-(\omega+\Delta)}{a} } ],~~  \omega+ \Delta < 0
     \\
	 B \Lambda ,~~ \omega+ \Delta > 0
    \end{array}     \right.
\label{dos}
\end{eqnarray}
   where $ \Lambda $ is the momentum cutoff and $ B=\frac{1}{ (2\pi)^2 b} $.
   Note the non-analytic depletion in the DOS due to the TPT.

   From the DOS, we can evaluate the ground state energy:
\begin{eqnarray}
    E \sim \left \{ \begin{array}{ll}
	 \alpha \Delta^2 + \frac{1}{ (2\pi)^2 b} \frac{2}{\sqrt{a} } \frac{2}{15} |\Delta|^{5/2} +  \cdots,~~  \Delta < 0
     \\
	 \alpha \Delta^2 + B_0 \Delta^3+ \cdots ,~~ \Delta > 0
    \end{array}     \right.
\label{energy}
\end{eqnarray}
   where $ \cdots $ means analytical terms or higher order non-analytic terms, the $ \alpha, B_0 $ are cutoff dependent.
   Only the leading non-analytic term is cutoff in-dependent and universal.
%   It is the $ \Delta $ dependence of $ b $  which leads to the background numerical value $ -0.77 $.

 Plugging the parameters $ \Delta=\sqrt{3}( \theta_c-\theta ), a=1/2, b=3/4 + \Delta/4 $
 into Eq.\ref{energy} and taking  two derivatives lead to:
\begin{eqnarray}
     E^{\prime \prime }(\theta, \mu=0 ) \sim \left \{ \begin{array}{ll}
	 \alpha + A_0 \sqrt{ \theta- \theta_c },~~~~\theta > \theta_c
     \\
	 \alpha + B_0 (\theta- \theta_c),~~~~\theta < \theta_c
    \end{array}     \right.
\label{energytwo}
\end{eqnarray}
   where the exponents $ \nu_{-}=1/2, \nu_{+}=1 $ are universal and the coefficient
   $ A_0=0.18856 $ is cutoff independent and stands for the universal contributions from a single cone in Fig.\ref{tqpt1}.
   While $  B_0 $ is not universal and cut-off dependent.
   At the half filling $ \mu =0 $, there are 6 particle and 6 hole WFS colliding at the same time.
   So in the second derivatives of the total ground state energy: $ A=12 A_0=2.262 $.

   We performed numerical calculations on the ground state energy in the BZ ( See SM ).
\begin{eqnarray}
 E^{\prime \prime }_n(\theta, \mu=0 ) \sim \left \{ \begin{array}{ll}
	-0.77 + A_n ( \theta- \theta_c )^{\nu_+},~~~~\theta > \theta_c
     \\
	-0.77 + B_n (\theta- \theta_c)^{\nu_-},~~~~\theta < \theta_c
    \end{array}     \right.
\label{energytwon}
\end{eqnarray}
   where the numerical exponents $ \nu_{+}=0.5\pm 0.05, \nu_{-}= 1.0\pm 0.05 $ match the analytical values $ \nu_{+}=1/2, \nu_{-}= 1 $ well,
   the numerical coefficient $ A_n =2.19 $ is also very close to the analytical value $ A=2.262 $ achieved above.

% In fact, the position of the Von Hove singularity ( saddle point ) also
% shift when away from the QCP, but it can be shifted away when calculating DOS and the ground state energy.

{\bf 4. Universal sub-leading scaling functions across the third TPT at a finite T. }
%\begin{figure}
%\includegraphics[width=8cm]{tqptkc.eps}
%\hspace{0.5cm}
%\includegraphics[width=3cm]{fmcritical.eps}
%\caption{ Topological quantum phase transition near a single collision momentum $ \vec{Q}_c=(Q_c,0,0) $. $ \Delta $ is the tuning parameter
%and $ Q=Q_c + \Delta/\sqrt{3} $ is the Von Hove singularity which coincides with $ Q_c $ only at the QCP $ \Delta=0 $. }
%\label{tqptcross}
%\end{figure}
% It was established \cite{scaling,sachdev} that near a quantum phase transition at zero temperature,
% the experimental measurable physical quantities such as single particle Green functions, specific heat,
% compressibility, magnetic susceptibilities, etc should satisfy scaling functions.
% However, there are always low energy excitations around the WFS on both sides of the TPT in Fig.\ref{typepp}.
% It becomes problematic to apply the scaling analysis near a quantum phase transition with an order parameter to such
% a TPT. Unfortunately, the previous RG analysis \cite{shankar,hightc2} designed to deal with leading scalings around an extended FS do not apply here
% due to the cone singularity of the FS geometry in Fig.\ref{typepp}.
 From Eq.\ref{TPTdis}, intuitively, one can still define the dynamic exponent $ z=2 $ with respect to the cone singularity.
 However, due to the low energy excitations around the WFS on both sides of the TPT, its physical meaning remains to be carefully examined.
% If it has any, it should be quite different from that defined in the QPT with an order parameter and symmetry breaking\cite{scaling,sachdev}
% and need to be carefully examined.
% Indeed we show that despite the leading terms in all these  physical quantities are cutoff dependent and non-universal,
 In the following, we show that it is the subleading terms which satisfy the universal scaling with $ z=2 $ and lead to
 non-analytic£¬ therefore non-Fermi liquid corrections to the leading analytic terms.
% They always take the opposite sign to the leading term, therefore can be called topological depletions.

\begin{figure}
\includegraphics[width=7cm]{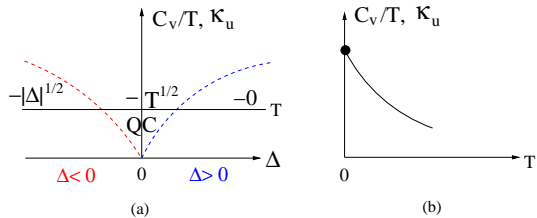}
\caption{ Experimental signatures of the topological depletions and sub-leading scalings
(a) The specific heat $ C_v/T $ and the compressibility  $ \kappa_u $ at a given $ T $ shows
 a universal non-analytic $ \sqrt{T} $ depletion in the quantum critical ( QC ) regime.
%no depletion when $ \Delta \gg T $ and a constant $ \sqrt{ |\Delta| } $ depletion when $ -\Delta \gg T $.
%The FS geometry in the three regimes are shown in Fig.\ref{typepp}.
%The crossover temperature is estimated $ T \sim \Delta \sim t \sim 3nK $ in cold atom experiment.
(b) The quantum $ \sqrt{T} $ cusps in  $ C_v/T $ and $ \kappa_u $ in the QC regime
as $ T $ lowers.  From the ratio of the coefficients of the $ \sqrt{T} $ in the two quantities,
one may also measure the universal Topological Wilson ratio $ R^{TD}_{W} $. }
\label{cusp}
%    It was established \cite{largecluster} that at the two $ SU(2) $ points at $ \theta=0 $ and $ \theta= \pi $,
%    there is a direct second order phase transition from the semi-metal phases to the AFM state at $ U/t \sim 3.8 $,
%    no intervening  gapped $ Z_2 $ spin liquid phase as claimed in \cite{meng}. However, it is not known if there is an intermediate $ Z_2 $ spin liquid
%    between the semi-metal phases and the $ Y-x $ stripy state away from the two $ SU(2) $ points.
\end{figure}

 One can apply the scaling analysis in \cite{tqpt} here to write down the sub-leading scaling functions
 for the specific heat and the uniform compressibility $  \kappa_u=\chi^{00}(\vec{q} \rightarrow 0, \omega=0 ) $
 for a single particle-particle ( or hole-hole ) cone in Fig.\ref{tqpt1}:
\begin{eqnarray}
  C_v & = & \frac{\pi^2}{3} B k_B (k_B T) \Lambda- \frac{ B k_B(k_B T)^{3/2} }{\sqrt{a}} \Psi_i( \frac{|\Delta|}{k_B T} )   \\  \nonumber
  \kappa_u & = & \frac{1}{2}B \Lambda -\frac{ B (k_B T)^{1/2} }{\sqrt{a}} \Omega_i( \frac{|\Delta|}{k_B T} )
\label{Tscaling}
\end{eqnarray}
   where $ i=1,2 $ stands for the two sides of the transitions $ \Delta < 0 $ and  $ \Delta > 0 $ in Fig.\ref{tqpt1} and Fig.\ref{cusp}.
   Note the first term is the leading term, proportional to the frequency ( or energy ) cutoff $ \Lambda $ and non-universal.
   while the second term is the sub-leading term, independent of the frequency ( or energy ) cutoff $ \Lambda $ and a
   universal function of the scaling variable $ s=\frac{|\Delta|}{k_B T} $. Due to the opposite sign between the two terms,
   the universal sub-leading term can be interpreted as the topological depletion coming from the TPT.

   The general form of the two scaling functions  $ \Psi_i(x) $ and $ \Omega_i(x) $ are evaluated in the SM.
   Here, we only list the topological depletions in the three limiting regimes in Fig.\ref{cusp} for the specific heat
\begin{align}
    C^{TD}_v=\begin{cases}
		-\dfrac{\pi^2}{3}\dfrac{B}{\sqrt{a}}k_B^2T\sqrt{|\Delta|},
		& \Delta\ll-k_BT\\
		-2.88201\dfrac{B}{\sqrt{a}}k_B^{5/2}T^{3/2},
		& |\Delta|\ll k_BT\\
		-\dfrac{\sqrt{\pi}}{2}
		\dfrac{Bk_B^{1/2}}{\sqrt{a}}\dfrac{\Delta^2}{\sqrt{T}}
		e^{-\frac{\Delta}{k_BT}},
		& \Delta\gg k_BT		
	\end{cases}
\label{spec}
\end{align}
    and for the uniform compressibility:
\begin{align}
    \kappa^{TD}_u= \begin{cases}
	    -\dfrac{B}{\sqrt{a}}\sqrt{|\Delta|},
	    &\Delta\ll-k_BT\\
	    -0.536077\dfrac{B k_B^{1/2}T^{1/2}}{\sqrt{a}},
	    &|\Delta|\ll k_BT\\
	    -\dfrac{\sqrt{\pi}}{2}\dfrac{B k_B^{1/2}T^{1/2}}{\sqrt{a}}
	    e^{-\frac{\Delta}{k_BT}},
	    &\Delta\gg k_BT
	\end{cases}
\label{comp}
\end{align}

  One can see both topological depletions are non-analytic only in the QC regime in Fig.\ref{cusp}.
  While, essentially no depletion when $ \Delta \gg T $ and a constant $ \sqrt{ |\Delta| } $ depletion when $ -\Delta \gg T $ which
  can be absorbed to the leading FL contribution anyway. This fact make their experimental detections feasible ( see Sec.7 ).

  One can also form the topological Wilson ratio
  $ R^{TD}_W\left(\frac{|\Delta|}{k_BT}\right) =\frac{k_B^2  T \kappa^{TD}_u}{C^{TD}_v} $ whose values in the three regimes are:
\begin{align}
    R^{TD}_W= \begin{cases}
	    3/\pi^2
	    &\Delta\ll-k_BT\\
	    0.186,
	    &|\Delta|\ll k_BT\\
	    \left(\frac{k_BT}{\Delta}\right)^2,
	    &\Delta\gg k_BT
	\end{cases}
\label{wilson}
\end{align}
   which is even independent of $ a $ and $ b $ characterizing the shape of the cone.
   In fact, it is also independent of how many cones are participating in the TPT, so universal
   for all the TPTs in Fig.\ref{tqpt1},\ref{tqpt2},\ref{tqpt3}.

   Due to the $ [C_4 \times C_4]_D $ symmetry at $ \alpha=\beta=\gamma $,
   the topological depletions of the  magnetic susceptibility $ \chi^{xx}(T)=\chi^{yy}(T)=\chi^{zz}(T)
    =\frac{1}{3}\chi^{00}(T)  $ also satisfy the sub-leading scaling Eq.\ref{comp}.

{\bf 5. The 3rd order TPT along  $ \alpha=\pi/2, \beta=\gamma=\theta $ and co-existence of four Type I Weyl fermions.  }
At half filling $ \mu=0 $, 2 particle WFS and 2 hole WFS collide at the same time at $ \vec{K}_c=(\pi/2,0,0 ) $
and $ \theta_c=\pi/3 $ ( Fig.\ref{tqpt2} ).
The dispersion near  $  K_c= \pm \pi/2, \theta_c=\pi/3 $ can also be written as Eq.\ref{TPTdis} where
$ \Delta=\sqrt{3}( \theta_c-\theta ), a=1/2, b=5/8 - \Delta^2/8 $.
Using Eq.\ref{energytwo}, we find $ A_0=0.2263 $, then the 2 particle WFS and 2 hole WFS contribute to $ A=4 A_0=0.90 $.
Similarly, the subleading scaling function in Eq.\ref{spec} and
Eq.\ref{comp} need also multiply by $ 4 $, but the topological Wilson ratio Eq.\ref{wilson} remains identical.
%Due to the non-conservation of the spins, only the sum over the spin components
%$ \sum_i \chi^{ii}(T)= \kappa_u $ in the magnetic susceptibilities
%satisfy the sub-leading scaling Eq.\ref{comp}.

  We also performed numerical calculation in the whole BZ and get a similar form as Eq.\ref{energytwon}
  where $ A_n=0.861 $ is quite close to the analytic value $ A=4 A_0=0.90 $.

 As shown in Fig.\ref{tqpt2}, there are also 4 Type I Weyl fermions  located at
 $ (0, 0, \pi), (0, \pi,0), (\pi,0,\pi), (\pi, \pi,0) $ with the anisotropic dispersion
 $ \epsilon^{I}_{\pm}= \pm \sqrt{ q^2_x+ \sin^2 \theta ( q^2_y+ q^2_z) } $.
 They reman intact through the TPT, so just act as 4 spectators.
 Due to its dynamic exponent $ z=1 $, their contributions $ C_v \sim T^{3}, \chi_u \sim T^{2} $
 are analytic and subleading to the topological non-analytic depletions in the QC regime  due to the third order TPT.

% From the simple scaling analysis with $ z=1 $,
% their contributions to the specific heat is $ C_v \sim T^{d/z} \sim T^3 $ which is analytic and subleading to the topological
% analytic depletion $ C_v \sim T^{d/z} \sim T^{3/2} $  in the QC regime  due to the third order TPT.
% Furthermore, it can not distinguished from the analytic $ T^3 $ FL corrections.
% However, they do contribute to the surface Fermi arcs and associated chiral anomalies in the transport properties.
% How the TPT in the bulk in Fig.\ref{tqpt2} interfere with the surface Fermi arcs need to be investigated  in a separate publication.

%Any $ U >0 $, we will determine how the orbital order pick up from one of the two
%P-H nesting vectors. Then it become a mixed phase of a spin-orbital correlated phase + 4 Weyl fermions.
%The mixed phase will undergo a quantum phase transition into some spin-orbital correlated phases described by
%the RAFHM Eqn.\ref{rh}.
%At $ U<0 $, gaining from the experiences from {\bf 2}, we will study the  transition from a possible Topological superconductor in Fig.10b
%to a trivial one in Fig.10d. Similarly, we will investigate possible topological Weyl Mott insulator transition at $ U>0 $.

{\bf 6. The 5th order TPT along the line $ \alpha=\beta=\pi/2, \gamma=\theta $ and the 8 Type II Weyl fermions.  }
% How the FS evolves along this line is shown in Fig.\ref{tqpt3}.
% As shown in Fig.\ref{tqpt3}, there is a TPT at the $ \pi $ flux ( in XY plane ) Abelian ending point $ \theta_c=0 $
% with the $ \tilde{\tilde{SU}}(2) $ symmetry in the rotated basis in Fig.\ref{cubicweyl}b.
 At half filling $ \mu=0 $, all the 4 particle WFS and 4 hole WFS collide at the same time at $ \theta_c=0 $ ( Fig.\ref{tqpt3} ).
Near $  \vec{K}_c=(\pi/2,0,0), \theta_c=0 $, the dispersion can also be written as Eq.\ref{TPTdis} where
$ \Delta=-\theta^2/2, a=1/2, b=1/2 + \Delta/2 $. Note the quadratic dependence of  $ \Delta $ on the SOC tuning parameter $ \theta $.
Plugging these parameters into Eq.\ref{energy}, we find the transition is a 5th order one with  $ \frac{ d^5 E}{d \theta^5} \sim \frac{8}{ \pi^2}  sgn \theta $.
Fig.\ref{tqpt3} dictates that $ A=12 A_0=\frac{96}{\pi^2} $.
All the subleading scaling function in Eq.\ref{spec} and Eq.\ref{comp} need also be multiplied by $ 12 $, but the topological Wilson ratio
Eq.\ref{wilson} remains the same.

%{\sl 1. Type II Weyl fermions at $ \alpha=\beta=\pi/2, \theta=0 $ as a TPT in particle-hole channel }
As shown in Fig.\ref{tqpt3}, in addition to the particle-particle and hole-hole WFS collisions,
the particle WFS also collide with the hole WFS at  the
8 momenta $ (0,0, \pm \pi/2), (\pi,0,\pm \pi/2),(\pi,\pi,\pm \pi/2), (0,\pi,\pm \pi/2) $. Such
a cone structure between the particle WFS and the hole WFS  is nothing but a special case of the type II Weyl fermions discussed in \cite{type2}.
%Shown in Fig.\ref{type2} is essentially a 3d version of 2d Dirac fermions.
%In 2d case, it was known \cite{rafhm} that there are 4 Dirac fermions at  $ \alpha=\beta=\pi/2 $ with topological charges
%$ 1,-1,1,-1 $ at the 4 Time-reversal invariant momenta $ (0,0), (\pi,0),(\pi,\pi), (0,\pi) $.
%Now adding the third dimension without putting any SOC along it will change the 4 Dirac fermions into the 8 Type II Weyl fermions
%at the 8 momenta $ (0,0, \pm \pi/2), (\pi,0,\pm \pi/2),(\pi,\pi,\pm \pi/2), (0,\pi,\pm \pi/2) $ shown in Fig.\ref{tqpt3}.
%Their topological charges are determined by the projections onto the $ ( k_x,k_y ) $ plane,
%independent of the $ k_z $ component, so still given by $ 1,-1,1,-1 $
%at the 4 projections on the $ ( k_x,k_y ) $ plane: $ (k_x,k_y)=(0,0), (\pi,0),(\pi,\pi), (0,\pi) $.
   One Type II Weyl fermion's dispersion at $ (0,0, \pi/2) $ is given by:
%\begin{equation}
 $ \epsilon^{II}_{\pm}(\vec{q})=-[ - q_z \mp \sqrt{ q^2_x + q^2_y } ] $
%\label{type2dis}
%\end{equation}
   where the $ \mp $ corresponds to the particle and hole WFS ( see Fig.\ref{tqpt3} ).
%   At $ \mu=0 $, taking $ - $ sign leads to the particle WFS $ -q_z \geq \sqrt{ q^2_x + q^2_y } $ which takes a cone structure near $ (0,0,\pi/2) $.
%   Taking $ + $ sign leads to the hole WFS $ q_z \geq \sqrt{ q^2_x + q^2_y } $ which also takes a cone structure
%   above the particle cone shown in Fig.\ref{type2}.
%   Now putting the SOC $ \gamma=\theta $ along the third direction, any small $ \theta $ immediately opens gap
%   to both the particle WFS $  -q_z \geq \sqrt{ \theta^2+ q^2_x + q^2_y } $ near $ (0,0,0) $ and
%   the hole WFS $ q_z \geq \sqrt{ \theta^2 + q^2_x + q^2_y } $ near $ (0,0,\pi) $ with the dynamic exponent $ z=1 $ in Fig.\ref{type2}.
%   At the same time, the 4 particle WFS split with the form Eq.\ref{TPTdis} with the dynamic exponent $ z=2 $.
%   The 4 hole WFS also split at the 4 FS nesting momenta.
%   When $ \theta $ gets close to $ \pi/2 $, then the particle and the hole WFS shrink to a small sphere
%   with $ q^2_{x}+ q^2_{y}+q^2_{z}= ( \pi/2-\theta)^2 $ shown in Fig.\ref{type1}.
%   At $ \theta=\pi/2 $, they shrink to Type I Weyl fermion  shown in Fig.\ref{cubicweyl}.
%   Although type I fermion is a semi-metal,
   In the Type II Weyl fermion, both the extended particle and hole WFS  add to contribute to a finite DOS
   $ D(\omega) \sim \Lambda^2- \alpha \omega^2 $ with a possible topological depletion in DOS $ \sim  \alpha \omega^2 $,
   so it is a metallic phase \cite{misleading}.
   For tilted Type II Weyl fermions in \cite{type2}, $ \alpha >0 $.
   While for the straight Type II Weyl fermions in Fig.\ref{tqpt3}, $ \alpha=0 $.
 From the simple scaling analysis, the 8 Type II Weyl fermions with $ z=1 $
 contribute to the depletion in the specific heat  $ C^{D}_v \sim \alpha T^3 $ which is subleading to the topological
 depletion $ C^{D}_v \sim  T^{3/2} $  due to the 5th order TPT in the QC regime. In fact,
 as said above, for the straight Type II Weyl fermions,
 even the coefficient $ \alpha=0 $,  so it does not even have a topological depletion.
%  $ C^{D}_v, \chi^{D}_{u} $.

{\bf 7. Experimental realizations and detections in cold atoms and materials. }
 The Hamiltonian Eq.\ref{ham} can be
 achieved by loading cold atoms in a cubic optical lattice \cite{expk40,expk40zeeman,2dsocbec,clock,clock1}.
 Note that the TPT in Fig.\ref{tqpt1},\ref{tqpt2},\ref{tqpt3}
are for non-interacting fermions, they are essentially single particle properties, so the heating issues should be manageable in current cold
atom experiments with fermions.
% Indeed, recently, 2d Rashba SOC has been experimentally implemented in the fermion $ ^{40} K $ gas \cite{expk40,expk40zeeman}.
% Soon after, using an optical Raman lattice scheme, the authors in the experiment \cite{2dsocbec}  realized the
% 2d Quantum anomalous Hall (QAH) SOC of spinor bosons with tunable anisotropy in a square lattice.
% Very recently, the optical lattice clock scheme \cite{clock}
%was implemented \cite{clock1,closk2,SDRB} to generate a 1d SOC in an optical lattice, it has the advantages to suppress the possible heatings issue.
% By using the most magnetic fermionic element dysprosium to eliminate the heating due to the spontaneous emission,
%the authors in \cite{ben} created a long-lived SOC gas of quantum degenerate atoms. The long lifetime of this weakly interacting SOC degenerate
%Fermi gas will facilitate the experimental study of quantum many-body phenomena manifest at longer time scales.

Of course, any experiments are performed at finite temperatures which are controlled by the
topological phase transitions at $ T=0 $ in Fig.\ref{tqpt1},\ref{tqpt2},\ref{tqpt3}.
%The TPTs do not survive at finite $ T >0 $, but become the 3 crossover regimes shown in Fig.\ref{cusp}.
The crossover temperature in Fig.\ref{cusp} can be estimated as $ T \sim \Delta \sim t \sim 3nK $
which is easily experimental reachable with the current cooling techniques \cite{cool1,cool2},
so the $ \sqrt{T} $ quantum cusp behaviors in $ C_v/T $ and $ \kappa_u $, the universal Topological Wilson ratio in Fig.\ref{cusp} can be
measured by the specific heat \cite{heat1,heat2},
{\sl In-Situ} \cite{dosexp} and the compressibility $ \kappa $ measurements \cite{heat2}.
The change of the FS topology across the TPT in Fig.\ref{tqpt1},\ref{tqpt2},\ref{tqpt3}
%and in the Type I Weyl fermions in Fig.\ref{type1} and Type II Weyl fermions in Fig.\ref{type2}
can be monitored by  the momentum resolved interband transitions \cite{topoexp}
and the band mapping technique developed in \cite{2dsocbec}.
%The heating issues with fermions may be more serious than those with spinor bosons.

%As shown in Fig.\ref{tqpt1},\ref{tqpt2},\ref{tqpt3}, Type I fermions are quite common and robust in this simple SOC model.
%However, the Type II fermions seem quite restricted, also only straight Type II fermions different than the tilted Type II
%proposed in \cite{type2} can be realized. Subleading scaling functions in Eq.\ref{spec} and \ref{comp} can be easily extended to the Type II Weyl fermions
%with $ z=1 $ in these materials. Unfortunately, the topological depletion $ \sim T^3 $ is analytic, can not be distinguished from the FL corrections
%which are also $ T^3 $. As said above, there is no such topological depletion for the  straight Type II fermions.
Type I Weyl fermions have been discovered in several materials \cite{weylexp1,weylexp3}.
Type II Weyl fermions \cite{type2} seem also have been found in a few materials \cite{type2arc2}.
%although there are still quite
%controversial experimental interpretations on the number of bulk type II Weyl fermions and associated surface Fermi arcs \cite{type2arc1,type2arc2}.
Topological Lifshitz transitions happen in all these Type I and Type II Weyl fermion materials.
Although they may not be described precisely by the Hamiltonian Eq.\ref{ham}, they should be in the same topological classes as
those in Fig.\ref{tqpt1},\ref{tqpt2},\ref{tqpt3}. So the results achieved in this paper should also apply to the Topological Lifshitz transitions
in these non-interacting or weakly interacting Type I and Type II Weyl fermion materials.

%In contrast to previous conventional quantum phase transitions \cite{scaling,sachdev}
%or topological phase transitions \cite{tqpt,dirac1,dirac3}, these TPT only contribute to sub-leading non-analytic
%topological depletions. From Eq.\ref{spec} and \ref{comp}, one can see that
%at a given $ T $, the topological depletions in $ C_v/T $ and $ \kappa_u $ reach maximum at the QC regime
%as shown in Fig.\ref{cusp}a.
%In the QC regime, as the $ T $ lowers, both $ C_v/T $ and $ \kappa_u $ show a $ \sqrt{T} $ quantum cusp at $ T=0 $ as shown in Fig.\ref{cusp}b.
%These sub-leading scaling behaviors could be detected

{\bf 8. Discussions and Conclusions. }
   Eq.\ref{spec} and \ref{comp} take a similar form to the topological entanglement entropy \cite{levin}: $ S= \alpha L - \gamma $
   where the first term ( Area law ) is the leading non-universal term  proportional to the length between the boundary of the two entangled
   regimes A and B. While the second term is the sub-leading term, independent of the boundary and universal called topological entanglement entropy
    $ \gamma= \log D $ where $ D $ is quantum dimension $ D $ ( which is a counter-part of the dynamic exponent $ z $ here ).
   There is also a relative minus sign between the two terms.
   This suggests that the form  may be a general scaling structure across a TPT.
%    in sharp contrast to
%   the conventional leading scaling across a conventional QPT with an order parameter and associated symmetry breaking.

  The quantum subleading scaling behaviors in the specific heat in Eq.\ref{spec}
  remind the classical cusp of the specific heat near the finite temperature phase transition of the  classical $ O(3) $ Heisenberg model \cite{3dheisen}
  $ C_v \sim C- b_0 t^{-\alpha} $ where $ t= |(T-T_c)/T_c |, b_0 >0  $ and $ \alpha \sim -0.1 $.
%  So the specific heat will show a  maximum classical cusp  near $ T_c $.
  This cusp has been precisely detected in specific heat experiments.
  This fact has also been used to determine the Anomalous Hall effect near the finite temperature phase transition in \cite{ahe}.
%  The $ O(3) $ heisenberg model may not be realized in the $ T=0 $ 2d Quantum Anti-ferromagnet due to the Berry phase effects.
  Here the quantum  $ \sqrt{T} $ cusp behavior in the QC regime near $ T=0 $ in Fig.\ref{cusp} is due to the TPT at $ ( T=0, \Delta=0 ) $.
%  It seems that the TPT discussed here is the first mechanism to show the quantum cusp behavior ( with a  constant background ).

  The present paper focused on only the half filling case with $ \mu=0 $. Our preliminary results away from half filling shows
  there are new classes of TPT with anisotropic dynamic exponents \cite{bosontype2}.
%  Of course, at sufficiently large $ \mu $,
%  there is a quadratic band touching through  a closed hole FS with the DOS $ D( \omega ) \sim \sqrt{ \omega }, \omega >0; D( \omega )=0, \omega < 0 $.
%  It leads a metal to BI transition which   satisfies a leading non-analytic scaling with dynamic exponent $ z=2 $.
%  The results will be presented in a subsequent publication.
%  It would be also interesting to look at the physical classification here from formal K theory classification.
%  However, so far, the K theory classification only assumes the translational symmetry, but ignore the constraints from crystalline symmetries
%  which take the spin-orbital coupled  crystalline symmetries in Eq.\ref{ham}.
  Due to the vanishing of DOS at the Type I Weyl fermions, a weak interaction is irrelevant. But due to
  the extended FS at the Type II Weyl fermions, the particle-particle WFS or hole-hole WFS TPT cone singularity,
  any weak interaction is relevant.
  Following Ref.\cite{rh,rhh,rhtran,rafhm,pairing,toposuper}, it is important to look at the effects of both positive $ U $ and negative $ U $.
%  For example, for $ U > 0 $ and away from half filling, due to the $ N_3=-2 $ in Fig.\ref{tqpt1},
%  depending on the signs of pairing amplitudes on different parts of WFS, it may lead to new TR invariant topological superfluids
%  with an associated Majornan surface mode \cite{zhang}.

  We thank Yu Yi-Xiang for the early participation of the project and acknowledge AFOSR FA9550-16-1-0412 for the supports.
  The work at KITP was supported by NSF PHY11-25915.


\begin{thebibliography}{99}
\bibitem{misleading} We think that the title "Type-II Weyl Semimetals " in Ref.\cite{type2} maybe mis-leading,
due to th extended particle and hole WFS, Type II Weyl fermions lead to a metal instead of a semi-metal.
\bibitem{bosontype2} The leading anisotropic dynamic exponents associated with the bosonic analog of the Type II Weyl fermions happen in
 the Rotated Ferromagnetic Heisenberg model (RFHM) in a longitudinal Zeeman field \cite{rhh}, but not in a transverse field \cite{rhtran}.
\end{thebibliography}
\end{document}